\documentclass[prd,aps,twocolumn,showpacs,nofootinbib,nobibnotes,superscriptaddress]{revtex4}
\usepackage{epsfig}
\usepackage{amssymb}
\usepackage{graphics}
\usepackage{bm}

\begin{document}

\title{Primordial magnetic field limits from cosmological data}

\date{\today~~KSUPT-10/2}

\author{Tina Kahniashvili}
\email{tinatin@phys.ksu.edu}
\affiliation{McWilliams Center for Cosmology and Department of
Physics, Carnegie Mellon University, 5000 Forbes Ave, Pittsburgh,
PA 15213, USA}
\affiliation{Department of Physics, Laurentian
University, Ramsey Lake Road, Sudbury, ON P3E 2C, Canada}
\affiliation{Abastumani Astrophysical Observatory, Ilia State
University, 2A Kazbegi Ave, Tbilisi, GE-0160, Georgia}

\author{Alexander G.\ Tevzadze}
\email{aleko@tevza.org} \affiliation{Abastumani Astrophysical
Observatory, Ilia State University, 2A Kazbegi Ave, Tbilisi,
GE-0160, Georgia}
\affiliation{Faculty of Exact and Natural
Sciences, Tbilisi State University, 1 Chavchavadze Ave, Tbilisi,
GE-0128, Georgia}

\author{Shiv K.\ Sethi}
\email{sksethi@andrew.cmu.edu} \affiliation{McWilliams Center for
Cosmology and Department of Physics, Carnegie Mellon University,
5000 Forbes Ave, Pittsburgh, PA 15213, USA}\affiliation{Raman
Research Institute, Sadashivanagar, Bangalore 560080, India }

\author{Kanhaiya Pandey}\affiliation{Raman
Research Institute, Sadashivanagar, Bangalore 560080, India }

\author{Bharat Ratra}
\email{ratra@phys.ksu.edu} \affiliation{Department of Physics,
Kansas State University, 116 Cardwell Hall, Manhattan, KS 66506, USA}


\begin{abstract}
We study limits on a primordial magnetic field arising from cosmological
data, including that from big bang nucleosynthesis, cosmic microwave background
polarization plane Faraday rotation limits, and large-scale
structure formation. We show that the physically-relevant
quantity is the value of the effective magnetic field, and
limits on it are independent of how the magnetic field was generated.
\end{abstract}


\maketitle

\section{Introduction}

There is much interest in the origin of the coherent part of the
large-scale $\mu$G magnetic fields in galaxies
\cite{Vallee04}.\footnote{On larger scales there have been two
recent claims of an observed lower limit of order $10^{-15}$ G on
the intergalactic magnetic field \cite{neronov,limit2}, as well
as one claimed detection of a field of this strength \cite{Ando}.
Prior to these observations, the intergalactic magnetic field was
observationally only limited to be smaller than a few nG.} A
leading possible explanation is that these large-scale magnetic
fields are the amplified remnants of a primordial seed magnetic
field generated in the early Universe \cite{Giovannini08,
Subramanian10,Kandus10}. Such early magnetogenesis models include
those in which seed magnetic field generation occurs during
inflation or shortly thereafter, or during a cosmological phase
transition (such as
the electroweak or QCD transition). Clearly the strength of the
seed magnetic field should be small enough so as to not generate
a larger than observed cosmological anisotropy. Magnetic field
energy density contributes to the relativistic (radiation) energy
density and thus another requirement is that it not exceed the
big bang nucleosynthesis (BBN) bound on the radiation energy
density.

There are two main questions that need to be answered: (l) Are
the amplitude and statistical properties of any of these seed
magnetic fields such that, after amplification by a realistic
model, they can explain the strengths and correlation lengths of
the observed magnetic fields in large-scale structures (LSSs)
such as galaxies? and, (2) Are any of the seed magnetic fields
detectable through cosmological observations, such as cosmic
microwave background (CMB) measurements\footnote{ A cosmological
magnetic field induces all three kinds of gravitational
perturbations, scalar, vector, and tensor; all three of which
contribute to CMB temperature and polarization anisotropies.} or
LSS observations? And, if yes, what are the observational
constraints on such primordial magnetic fields?

In this paper we focus on the second question and consider two
cosmological consequences of a primordial magnetic field, Faraday
rotation of the CMB polarization plane and the effect on LSS
formation. As two of the effects of a primordial magnetic field,
these have been widely discussed in the literature. See Refs.\
\cite{cmb-b,sb,jedamzik98,cmb,mkk02,kr05} for discussions of
magnetic field induced CMB anisotropies,\footnote{The effects of a
homogeneous magnetic field on the scalar mode of CMB fluctuations,
including the resulting non-Gaussianity of the CMB temperature map,
are discussed in Refs. \cite{cmb0}.} Refs.\ \cite{kl,faraday,kklr05}
for the Faraday rotation effect, and Refs.\
\cite{LSS0,gs03,LSS,sb05,s08} for effects of a primordial magnetic
field on LSS formation.

It has become conventional to derive the cosmological effects
of a seed magnetic field by using a  magnetic field spectral shape
(parametrized by the spectral index $n_B$) and the smoothed value
of the magnetic field ($B_\lambda$) at a given scale $\lambda$
(which is usually taken to be 1 Mpc). We
develop here a different and more correct formalism based on the
effective magnetic field value that is determined by the total
energy density of the magnetic field. As a striking consequence,
we show that even an extremely small smoothed magnetic field of
$10^{-29}$ G at 1 Mpc, with the Batchelor spectral shape
($n_B=2$) at large scales, can leave detectable signatures in CMB
or LSS statistics.\footnote{This strong limit on the primordial
magnetic field is the consequence of the BBN bound and the sharp
shape of the magnetic field at large scales
\cite{caprini09,ktr09}. The low efficiency of gravitational wave
production by a cosmological magnetic field \cite{ktr09} results
in a weaker limit on the seed field from the direct detection of
the induced gravitational wave signal \cite{caprini01}.}

We also show that the conventional approach based on the smoothed
magnetic field results in some confusion when considering
phase-transition generated magnetic fields \cite{pmf} with spectral
shape sharper (on large scales) than the white noise spectrum (i.e.\
with $n_B > 0$). In this case the total energy density of the
magnetic field, $\rho_B$, is mainly concentrated on large
wavenumbers (small length scales). In what follows we show that
primordial magnetic field effects on cosmological scales are
determined not by the amplitude of the magnetic field on these
scales, but rather by the total energy density of the magnetic
field.

Here we consider limits on a seed magnetic field from the
observational constraint on the CMB polarization plane rotation
angle and observational constraints on the formation of the first
bound structures. Both tests give comparable limits on the
effective magnetic field, ranging from $10^{-9}$ G to
$10^{-7}$ G, depending on the spectral shape of the magnetic field.
Note that these limits are of order of the BBN bound. The best
limit on the seed magnetic field is for the scale-invariant case
that can be generated during inflation
\cite{ratra92,bamba}.

In a Universe with only scalar mode perturbations the CMB
$B$-polarization signal vanishes. The CMB $B$ polarization signal
can arise from vector or tensor perturbations, and thus
$B$-polarization detection based tests are powerful tools for
probing non-standard cosmological models and the relic gravitational
wave background. Since a cosmological magnetic field can source a
CMB $B$-polarization signal, a crucial test to limit the magnetic
field is based on CMB $B$-polarization measurements. In a separate
paper we will address the cosmological magnetic field energy density
limits that can result from this test.

The structure of our paper is as follows. In Sec.\ II we review
magnetic field statistical properties. In Sec.\ III we examine the
CMB polarization plane Faraday rotation effect and the resulting
magnetic field limits. In Sec.\ IV we consider the influence of a
magnetic field on LSS statistics and determine the resulting limits
on the magnetic field. We discuss our results and conclude in Sec.\
V.

\section{Modeling the primordial magnetic field}

A stochastic Gaussian magnetic field is fully described by its
two-point correlation function. For simplicity we consider here the
case of the non-helical magnetic field for which the two-point
correlation function in wavenumber space is \cite{mkk02}
 \begin{equation}
 \langle B^\star_i({\mathbf k})B_j({\mathbf k'})\rangle
=(2\pi)^3 \delta^{(3)} ({\mathbf k}-{\mathbf k'})
P_{ij}({\mathbf{\hat k}}) P_B(k). \label{spectrum}
\end{equation}
Here $i$ and $j$ are spatial indices, $i,j \in (1,2,3)$,
$\hat{k}_i=k_i/k$  a unit wavevector, $P_{ij}({\mathbf{\hat
k}})=\delta_{ij}-\hat{k}_i\hat{k}_j$
 the transverse plane projector,
 $\delta^{(3)}({\mathbf k}-{\mathbf k'})$  the Dirac delta
 function, and $P_B(k)$ is the power spectrum of the magnetic
field.\footnote{We use
\begin{equation}
   B_j({\mathbf k}) = \int d^3\!x\,
   e^{i{\mathbf k}\cdot {\mathbf x}} B_j({\mathbf x}),\ \ \
   B_j({\mathbf x}) = \int {d^3\!k \over (2\pi)^3}
   e^{-i{\mathbf k}\cdot {\mathbf x}} B_j({\mathbf k}), \nonumber
\end{equation}
when Fourier transforming between position and wavenumber spaces.
We assume flat spatial hypersurfaces (consistent with current
observational indications, \cite{rv08}).}

We define the smoothed magnetic field $B_\lambda$ through the
mean-square magnetic field \cite{mkk02}
\begin{equation}
{B_\lambda}^2 = \langle {\mathbf B}({\mathbf x}) \cdot {\mathbf
B}({\mathbf x})\rangle |_\lambda,
\end{equation}
where the smoothing is done on a comoving length $\lambda$ with a
Gaussian smoothing kernel function $\propto
\mbox{exp}[-x^2/\lambda^2]$. Corresponding to the smoothing length
$\lambda$ is the smoothing wavenumber $k_\lambda=2\pi/\lambda$. The
power spectrum $P_B(k)$ is assumed to be a simple power law on large
scales, $k<k_D$ (where $k_D$ is the cutoff wavenumber),
\begin{equation}
P_B(k) = P_{B0}k^{n_B}= \frac{2\pi^2 \lambda^3
B^2_\lambda}{\Gamma(n_B/2+3/2)} (\lambda k)^{n_B},
\label{energy-spectrum-H}
\end{equation}
and assumed to vanish on small scales where $k>k_D$.

The energy density of the magnetic field is \cite{kr05}
\begin{equation}
  {\rho}_B(\eta_0)
= \frac{B_\lambda^2 (k_D \lambda)^{n_B+3}}{8\pi
\Gamma(n_B/2+5/2)}.
\end{equation}
We define the effective magnetic field $B_{\rm eff}$ through $\rho_B
= {B_{\rm eff}^2}/({8\pi})$  and thus we get for the scale-invariant
spectrum ($n_B = -3$ \cite{ratra92}) $B_{\rm eff} = B_\lambda$ for
all values of $\lambda$. The scale-invariant case is the only case
where the values of the effective and smoothed fields coincide.

We need to define the magnetic field cut-off wavenumber $k_D$. We
assume that the cut-off scale is determined by the Alfv\'en wave
damping scale $k_D \sim v_A L_S$ where $v_A$ is the Alfv\'en
velocity and $L_S$ the Silk damping scale \cite{jedamzik98}. Such a
description is more appropriate when we are dealing with an
homogeneous magnetic field and the Alfv\'en waves are the
fluctuations ${\bf B}_1({\bf x})$ with respect to a background
homogeneous magnetic field ${\bf B}_0$ ($|{\bf B}_1 |\ll |{\bf
B_0}|$). In the case of the stochastic magnetic field we generalize
the Alfv\'en velocity definition, see Ref. \cite{mkk02}, by
referring to the analogy between the effective magnetic field and
the homogeneous magnetic field. Assuming that the Alfv\'en velocity
is determined by $B_{\rm eff}$, a simple computation gives the
expression of $k_D$ in terms of $B_{\rm eff}$ \cite{ktr09}:
\begin{equation}
\frac{k_D}{1{\rm Mpc}^{-1}} = 1.4 \sqrt{\frac{(2\pi)^{n_B+3}
h}{\Gamma(n_B/2+5/2)}} \left(\frac{10^{-7}{\rm
G}}{B_{\rm eff}}\right). \label{rho1}
\end{equation}
Here $h$ is the Hubble constant in units of 100 km s$^{-1}$
Mpc$^{-1}$. The BBN limit on the effective magnetic field strength,
$B_{\rm eff} \leq 8.4 \times 10^{-7}$ G \cite{ktr09}, gives an upper
limit on the cut-off wavenumber $k_D$,
\begin{equation}
k_D^{\rm BBN} \geq 0.17 h^{1/2}  \frac{(2\pi)^{(n_B+3)/2}}{
\Gamma^{1/2}(n_B/2+5/2)}~{\rm Mpc}^{-1}.
\end{equation}
In the case of an extremely large magnetic field it is possible to
have $\lambda_D > 1$ Mpc. At this point it would seem unreasonable
(unjustified) to consider a smoothing scale $\lambda=1$ Mpc as is
conventionally done.

\section{CMB polarization plane rotation}

The presence of a primordial magnetic field during recombination
causes a rotation of the CMB polarization plane through the Faraday
effect \cite{kl}. The rms rotation angle $\alpha_{\rm rms} =
(\langle {\alpha }^2 \rangle)^{1/2}$ induced by a stochastic
magnetic field with smoothed amplitude $B_\lambda$ and spectral
index $n_B$ is given by
\begin{equation}
   \langle {\alpha }^2\rangle
   = \sum_{l} \frac{2l+1}{4\pi} C_l^{\alpha},
\end{equation}
where the rotation multipole power spectrum $C_l^\alpha $ is \cite{kklr05}
\begin{eqnarray} \label{ClRR-sym-int}
   & & C_l^\alpha \simeq  \\
   & & \frac{9 l(l+1)}{(4\pi)^3 q^2 \nu_0^4}
   \frac{B^2_\lambda}{\Gamma\left(n_B/2 +3/2 \right)}
   \left(\frac{\lambda}{\eta_0}\right)^{n_B+3}
   \!\!\!\! \int_0^{x_S} \!\! dx \, x^{n_B}j^2_l(x). \nonumber
\end{eqnarray}
Here $\eta_0$ is the conformal time today, $\nu_0 $ is the CMB
photon frequency, $q^2 = 1/137$ is the squared elementary charge in
cgs units, $j_l(x)$ is a Bessel function with argument $x=k\eta_0$,
and $x_S =k_S \eta_0$ where $k_S = 2~{\rm Mpc}^{-1}$ is the Silk
damping scale. In the case of an extreme magnetic field which just
satisfies the BBN bound, $k_D $ might become less than the Silk
damping scale. In this case the upper limit in the integral above
must be replaced by $x_D = k_D \eta_0$.

In terms of $B_{\rm eff}$, Eq.\ (\ref{ClRR-sym-int}) can be
rewritten in the following form,
\begin{eqnarray}
C_l^\alpha &\simeq & 1.6  \times 10^{-4} \frac{l(l+1)}{ (k_D
\eta_0)^{n_B +3}} \left( \frac{B_{\rm eff}}{1\, {\rm
nG}}\right)^2 \left( \frac{100 \, {\rm GHz}}{\nu_0} \right)^4
\nonumber \\ && \times \frac{n_B +3}{2} \int^{x_S}_0 dx \,
x^{n_B}j^2_l(x) ,
\end{eqnarray}
and, as a result,
\begin{eqnarray}
\alpha_{\rm rms} & \simeq & 0.14^\circ
\left(\frac{B_{\rm eff}}{1\, {\rm nG}}\right)   \left(
\frac{100 \, {\rm GHz}}{\nu_0} \right)^2 \frac{\sqrt{n_B +3}}{ (k_D
\eta_0)^{(n_B +3)/2}} \nonumber \\
& \times & \left[ \sum_{l=0}^{\infty} (2l+1)
  l(l+1) \int^{x_S}_0 dx \, x^{n_B}j^2_l(x) \right]^{1/2}\!\!\!.
\label{al}
\end{eqnarray}
It is of interest to compare Eq.\ (\ref{al}) with the corresponding
result, Eq.\ (2) of Ref. \cite{kl}, derived for a homogeneous
magnetic field and at frequency $\nu_0 = 30 $ GHz,
\begin{equation}
\alpha_{\rm rms} \simeq 1.6^\circ \left(
\frac{B_0}{1\, {\rm nG}}\right) \left( \frac{30\, {\rm GHz}}{\nu_0}
\right)^2 \label{al0}
\end{equation}
Both expressions agree for $n_B \rightarrow -3$ after
accounting for $\sum_l (2l+1) j_l^2 (x) =1$ and the fact that
Bessel functions peak at $x \sim l$ for given $l$ (see
Appendix A).

\begin{figure}
\begin{center}
\includegraphics[width=\columnwidth]{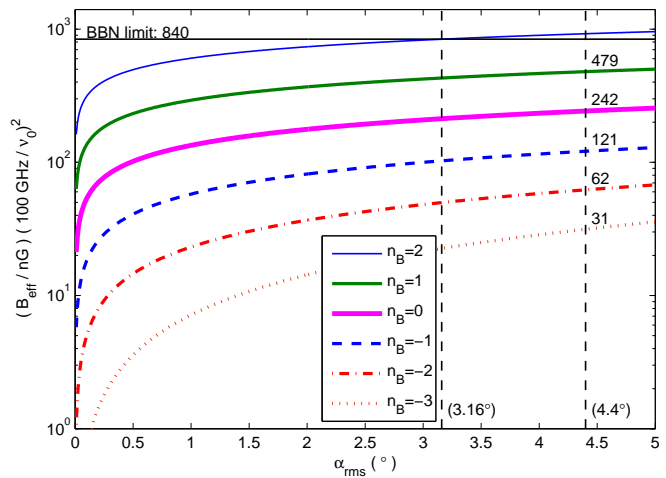}
\end{center}
\caption{Effective magnetic field limits set by the measurement
of the rotation angle $\alpha_{\rm rms}$ for different spectral
indices ($n_B = -3,-2,-1,0,1,2$, from bottom to top). The
horizontal solid line shows the upper limit set by BBN. Vertical
dashed lines correspond to the angles $\alpha_{\rm
rms}=3.16^\circ$ that is set by the BBN limit on the effective
magnetic field with spectral index $n_B=2$ and $\alpha_{\rm
rms}=4.4^\circ$ set by the WMAP 7-year data. The numerical values
of the effective magnetic field constraints (in nG at 100 GHz)
from the $\alpha_{\rm rms}=4.4^\circ$ limit are shown on the
graph for each spectral index value.} \label{Beff_lim}
\end{figure}

 Figure \ref{alpha_n} shows the rms rotation angle $\alpha_{\rm
rms}$, Eq.\ (\ref{al}), as a function of the spectral index $n_B$ when
the effective magnetic field is normalized to be $10^{-9}$ G.
The WMAP 7-year data limits the rms rotation angle to be less then
$4.4^\circ$ at $95\%$ C. L.\ \cite{WMAP7}. This allows us to limit
the effective magnetic field as shown in Fig.\ \ref{Beff_lim}.

\begin{figure}
\begin{center}
\includegraphics[width=\columnwidth]{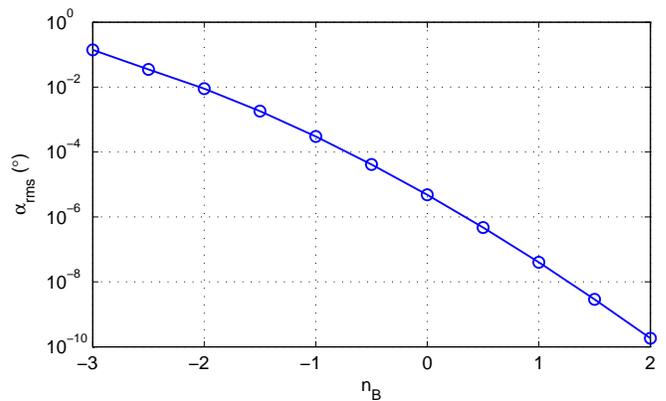}
\end{center}
\caption{Rms rotation angle $\alpha_{\rm rms}$ as a function
of spectral index $n_B$ for the case when $B_{\rm
eff}=1\, {\rm nG}$ and $\nu_0 = 100\, {\rm GHz}$. Circles
correspond to the computed values.} \label{alpha_n}
\end{figure}

\section{Large-scale structure}

A primordial tangled magnetic field can also induce the formation of
structures in the Universe. In particular, these fields can play an
important role in the formation of first structures (see, e.g.\
Refs.\ \cite{wasserman,kim,sb,gs03,sb05,s08,LSS0,LSS}).

The magnetic-field-induced matter power spec\-trum $P(k) \ {\rm is}
\propto k^4$ for $n_B > -1.5$ and $\propto k^{2n_B+7}$ for $n_B \le
-1.5$ \cite{kim,gs03}. The cut-off scale of the power spectrum  is
determined  by the larger of the magnetic Jeans' wavenumber $k_{\rm
J}$ and the thermal Jeans' wavenumber $k_{\rm therm}$ (for a
detailed discussion, see, e.g.\ Ref.\ \cite{s08}). Here the magnetic
Jeans' wavenumber is (see, e.g.\ Ref.\ \cite{kim})
\begin{equation}
k_J  \simeq  (230^{(n_B+3)/2}\times 13.8)^{2/(n_B+5)}\left({1 \, {\rm
nG} \over B_{\rm eff}} \right ) \, \rm Mpc^{-1}.
\end{equation}
Unlike the $\Lambda \rm CDM$ matter power spectrum, the
magnetic-field-induced matter power spectrum increases at small
scales and can exceed the $\Lambda \rm CDM$ matter one at small
scales (for a comparison of these two spectra, see, e.g.\ Fig.\ 3 of
Ref.\ \cite{gs03}). And, therefore, one of the more important
contributions of the additional power induced  by magnetic fields is
to the formation of the first structures in the Universe (e.g.\
Refs.\ \cite{sb05,LSS,LSS0} and references therein).

\begin{figure}
\epsfig{file=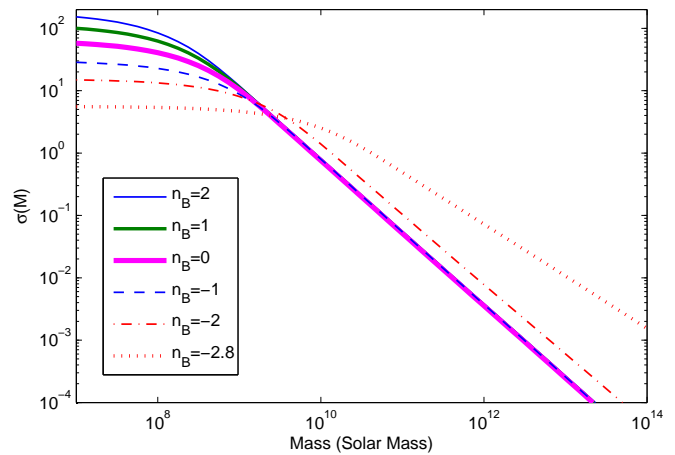,width=\columnwidth}
\caption{ The mass dispersion at $z = 10$ for $B_{\rm eff} =
6 \, \rm nG$ as a function of magnetic field power spectral
index $n_B$. From top to bottom (at the left hand side of the
plot), the curves correspond to $n_B = 2,1,0,-1,-2,-2.8$.}
\label{fig:f1}
\end{figure}

In Fig.\ \ref{fig:f1} we show the linear mass dispersion $\sigma(M)$
for matter power spectra induced by a primordial magnetic field with
$B_{\rm eff} = 6 \, {\rm nG}$ at $z = 10$ for different values of
$n_B$.  Notable features of Fig.\ \ref{fig:f1} are: (a) the mass
dispersion on small scales is larger for a larger value of $n_B$;
and, (b) for $n_B \ge -1.5$, the mass dispersion drops more sharply
at larger scales  than for $n_B \le -1.5$. We focus here on the mass
dispersion on the smallest scales, as these scales are more relevant
for the formation of the first structures in the Universe. These
first structures were responsible for the  reionization of the
Universe at $z \simeq 10$. To obtain meaningful constraints on
$B_{\rm eff}$ from the formation of first structures, we need to
know how the curves shown in Fig.\ \ref{fig:f1} vary as $B_{\rm
eff}$ is changed and as the Universe evolves.

The mass dispersion $\sigma(M,z)$ evolves with the time dependence
of the growing mode of the linear density perturbations sourced by
the primordial magnetic field \cite{kim,gs03}. The growing mode is
$\propto a(t)$, the scale factor, at high redshifts, the same as in
the ``standard'' $\Lambda$CDM case without a magnetic field. To
account for this evolution the curves corresponding to $\sigma$ in
Fig.\ \ref{fig:f1} must be scaled by roughly a factor of $\simeq
11/(1+z)$ for redshifts $z \gg 1$.

It can be shown that the value of $\sigma$ at the smallest scales
($M \simeq 10^6 \, \rm M_\odot$) is invariant under a change in
$B_{\rm eff}$ if the cut-off scale is determined by $k_{\rm J}$:
an  increase/decrease in the value of $B_{\rm eff}$ is
compensated by a decrease/increase in the value of $k_{\rm J}$.
However, if $B_{\rm eff}$ is decreased to a value at which
$k_{\rm therm} \le k_{\rm J}$, then the value of $\sigma$
decreases with a decrease in $B_{\rm eff}$, as the cut-off scale
becomes independent of the value of $B_{\rm eff}$.

It has been shown that the dissipation of magnetic fields in the
post-recombination era can substantially alter the thermal and
ionization history of the universe \cite{LSS0,sb05,s08}. In
particular, this dissipation raises the matter temperature and
therefore the Jeans' scale in the IGM. For $B_{\rm eff} \ge 1\, \rm
nG$ the matter temperature rises to $\simeq 10^{4} \, \rm K$ as
early as $z \ge 100$, \cite{sb05}, resulting in a steep rise in the
Jeans' scale as compared to the usual case. The Jeans' wave number
corresponding to this temperature is $k_{\rm therm} \simeq 10 \, \rm
Mpc^{-1}$ (see, e.g.\ Fig.\ 4 of Ref.\ \cite{s08}).

\begin{figure}
\epsfig{file=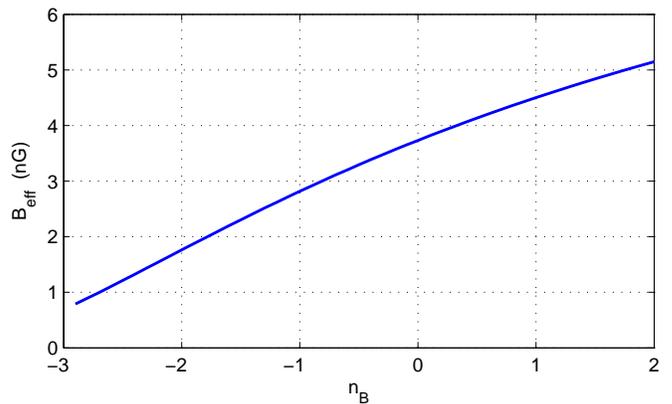,width=\columnwidth}
\caption{Constraint on the magnetic field strength $B_{\rm eff}$
as a function of the power spectral index $n_B$.}
\label{fig:f2}
\end{figure}

WMAP results show that the Universe reionized at $z \simeq 10$. This
reionization was caused by the non-linear collapse of the first
structures, followed by star formation  and the emission of UV
photons from the collapsed halos. For a virialized structure in the
spherical collapse model, the linear mass dispersion $\sigma \simeq
1.7$. This implies that the value of $\sigma$ at the scales of
interest at $z \simeq 10$ is not expected to be much higher than
$1.7$. Consider the $n_B = 2$ model in Fig.\ \ref{fig:f1}; the value
of mass dispersion at the smallest scales is $\simeq 100$, which
means that the first structures formed at $z \simeq 650$ in this
case (the redshift of the collapse of  first structures is $\simeq
6.5 \sigma_{\rm max}$, where $\sigma_{\rm max}$ is the maximum value
of $\sigma$ at $z \simeq 10$), which can certainly be ruled out by
the WMAP data on CMB anisotropies. A similar arguments can be used
to rule out almost all the models shown in Fig.\ \ref{fig:f1}. Only
the nearly scale-invariant models with $n_B \simeq -3$ do not put
strong constraints on the strength of the magnetic field. As argued
above, the value of mass dispersion at the smallest scales to
collapse is nearly independent  of the magnetic  field strength
unless $B_{\rm eff}$ decreases to a value such that $k_{\rm J} =
k_{\rm therm}$. In this case, the value of $\sigma$ decreases below
those shown in Fig.\ \ref{fig:f1}. We have explored a wide range of
$B_{\rm eff}$ for the range of spectral indices shown in Fig.\
\ref{fig:f1}. We find that the range of acceptable values is
$1\hbox{--}3 \, \rm nG$. In Fig.\ \ref{fig:f2} we show the $B_{\rm
eff}$ corresponding to $k_{\rm J} = k_{\rm therm}$. Notwithstanding
various complications discussed above, this figure gives a rough
sense of the acceptable range of $B_{\rm eff}$ over the entire range
of $n_B$.

In the foregoing, we neglect the impact of the $\Lambda$CDM model on
the process of reionization. As the density fields induced by the
$\Lambda$CDM model and the magnetic field are uncorrelated, the
matter power spectra owing to these two physical phenomena would add
in quadrature. The smallest structures to collapse at $z \simeq 10$
in the WMAP-normalized $\Lambda$CDM model are 2.5$\sigma$
fluctuations of the density field as opposed to the magnetic field
case where 1$\sigma$ collapse is possible (Fig.\ 3). This means  the
number of collapsed halos is more abundant in the latter case.
Therefore, depending on the star-formation history, if the
magnetic-field-induced halo collapse made an important contribution
to the reionization process, the far rarer halos from  $\Lambda$CDM
would have made a negligible impact (for further details and
references see Ref.\ \cite{LSS}).

\section{Conclusions}

In this paper we study the large-scale imprints of a cosmological
magnetic field, such as the rotation of the CMB polarization plane
and formation of the first bound structures. We derive the
corresponding limits on a primordial magnetic field energy density,
expressed as limits on the effective value of the magnetic field,
$B_{\rm eff}$. These limits are identical to limits on the smoothed
magnetic field $B_\lambda$ (smoothed over a length scale $\lambda$
that is conventionally taken to be 1 Mpc) only in the case of the
scale-invariant magnetic field (when $n_B = -3$). For a steep
magnetic field with spectral index $n_B =2$ the difference between
$B_{\lambda = 1\ {\rm Mpc}} $ and $B_{\rm eff}$ is enormous (greater
than $10^{15}$). We show that using the smoothed magnetic field can
result in some confusion; e.g.\ an extremely small smoothed magnetic
field on large scales does not mean that this field cannot leave
observable traces on cosmological scales.

An intergalactic magnetic field of effective value larger than 1-10
nG (with, depending on magnetic spectral index, corresponding values
of $B_{\lambda=1\ {\rm Mpc}}$ in the range $10^{-8}-10^{-26}$ G) is
ruled out by cosmological data. These limits of 1-10 nG are
consistent with recent observational bounds on the intergalactic
magnetic field \cite{neronov,limit2,Ando} if the field was
generated in the early Universe with spectral shape $n_B \leq 1$.
This favors the inflationary magnetogenesis scenario.

\acknowledgments

We acknowledge partial support from Georgian National Science
Foundation grant GNSF ST08/4-422, Department of Energy grant DOE
DE-FG03-99EP41043, Swiss National Science Foundation SCOPES grant
no. 128040, and NASA Astrophysics Theory Program grant NNXlOAC85G.
T.K.\ acknowledges the ICTP associate membership program.

\begin{appendix}

\section{Evaluating the right hand side of Eq.\ (\ref{al}) when $n_B
\rightarrow -3$}

\begin{figure}
\includegraphics[width=\columnwidth]{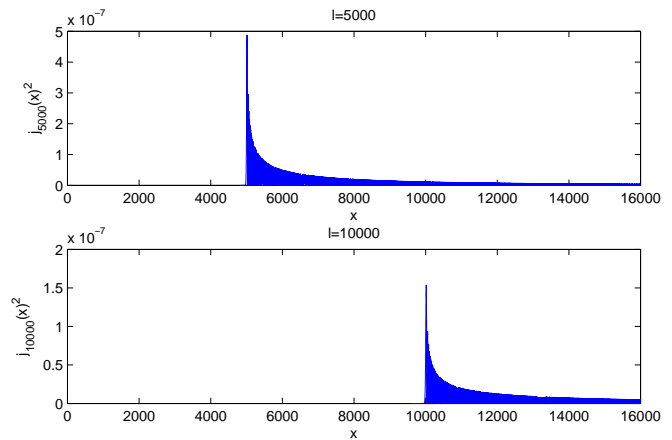}
\caption{The squared spherical Bessel functions $j_l^2(x)$ for
$l=5000$ (top) and $l=10000$ (bottom). Clearly $j_l^2(x)$ peaks
at $x \approx l$.} \label{Bes1}
\end{figure}
\begin{figure}
\includegraphics[width=\columnwidth]{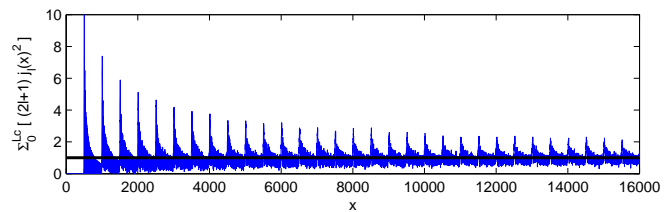}
\caption{The sum of the squared spherical Bessel function $
\sum_{l=0}^{l_C} (2l+1) j_l^2(x)$ for $l_C = x_S \simeq 16000$. The
sum converges to 1 (horizontal solid line).}
\label{Bes2}
\end{figure}

The $\sqrt{n_B + 3}$ factor in the numerator of the right hand side
of Eq.\ (\ref{al}) is compensated by a corresponding $1/\sqrt{n_B +
3}$ from the Bessel function integral when the spectral index $n_B
\rightarrow -3$ and so the expression for $\alpha_{\rm rms}$ remains
finite in this limit. To establish this we use properties of the
Bessel function. Recall that $j_l^2(x)$ peaks at $x\sim l$, as shown
in Fig.\ \ref{Bes1}. This allows us to replace the factor $l(l+1)
j_l^2(x)$ by $x^2 j_l^2(x)$ (the accuracy of this approximation is
of order 15-20$\%$). The next step is to perform the sum over $l$.
It is obvious that there is cut-off multipole number $l_C$ that
corresponds to the cut-off wavenumber, $l_C \sim {\rm min} (x_D,
x_S)$. Now $j_l^2(x)$ satisfies
\begin{equation}
\sum_{l=0}^{\infty} (2l+1) j_l^2(x)=1,
\end{equation}
while we are interested in computing $\sum_{l=0}^{l_C}(2l+1) j_l^2(x)$.
The Silk damping scale cutoff multipole number is $l_S \simeq 16000$,
\cite{kklr05}. Figure \ref{Bes2} shows that the sum to $l_S$ converges
to 1.

\end{appendix}

\end{document}